\documentclass[twocolumn,showpacs,preprintnumbers,amsmath,amssymb,reprint,prl,aps,superscriptaddress]{revtex4-1}
\pdfoutput=1
\usepackage{graphicx}
\usepackage{amsmath}
\usepackage{verbatim}
\usepackage{hyperref}

\newcommand{\QA}{\mathrm{Q}_{a}}
\newcommand{\QB}{\mathrm{Q}_{b}}
\newcommand{\QK}{\mathrm{Q}_{k}}

\newcommand{\wq}{\omega_{01}}
\newcommand{\wl}{\omega_{12}}
\newcommand{\wA}{\omega_{01}^a}
\newcommand{\wB}{\omega_{01}^b}

\newcommand{\wtA}{\omega_{12}^a}
\newcommand{\wtB}{\omega_{12}^b}

\newcommand{\wm}{\omega_m}

\newcommand{\tg}{t_{\mathrm{g}}}
\newcommand{\ts}{t_{\mathrm{s}}}
\newcommand{\tb}{t_{\mathrm{b}}}
\newcommand{\ti}{t_{\mathrm{i}}}

\newcommand{\Am}{A_m}
\newcommand{\Api}{A_{\pi}}

\newcommand{\EPG}{\mathrm{EPS}}

\newcommand{\ns}{\mathrm{ns}}

\newcommand{\MHz}{\mathrm{MHz}}
\newcommand{\GHz}{\mathrm{GHz}}

\newcommand{\ket}[1]{\left\lvert #1 \right\rangle}

\newcommand{\ketB}[1]{\left\lvert #1_b \right\rangle}
\newcommand{\ketK}[1]{\left\lvert #1_k \right\rangle}

\begin{document}
\title{Mitigating information leakage in a crowded spectrum of weakly anharmonic qubits}
\author{V.~Vesterinen}
\affiliation{VTT Technical Research Centre of Finland, P.O. Box 1000, 02044 VTT, Finland}
\affiliation{Kavli Institute of Nanoscience, Delft University of Technology, P.O. Box 5046,
2600 GA Delft, The Netherlands}
\author{O.-P.~Saira}
\affiliation{Kavli Institute of Nanoscience, Delft University of Technology, P.O. Box 5046,
2600 GA Delft, The Netherlands}
\author{A.~Bruno}
\affiliation{Kavli Institute of Nanoscience, Delft University of Technology, P.O. Box 5046,
2600 GA Delft, The Netherlands}
\author{L.~DiCarlo}
\affiliation{Kavli Institute of Nanoscience, Delft University of Technology, P.O. Box 5046,
2600 GA Delft, The Netherlands}
\date{\today}

\begin{abstract}
A challenge for scaling up quantum processors using frequency-crowded, weakly anharmonic qubits is to drive individual qubits without causing leakage into non-computational levels of the others, while also minimizing the number of control lines.  To address this, we  implement single-qubit Wah-Wah control in a circuit QED processor with a single feedline for all transmon qubits, operating at the maximum gate speed achievable given the frequency crowding.  Randomized benchmarking and quantum process tomography confirm alternating qubit control with $\leq1\%$ average error per computational step and decoherence-limited idling of one qubit while driving another with a Wah-Wah pulse train.
\end{abstract}

\maketitle

Experimental quantum computing~\cite{Ladd10} seldom employs true qubits. Most architectures use effective qubits defined by a pair of energy levels within a multi-level quantum object (typically the ground and first excited states, labelled $\ket{0}$ and $\ket{1}$). Examples include non-spin-1/2 electron and nuclear spins~\cite{Hanson08}, electronic levels in atoms and ions~\cite{Monroe13}, photons with combined polarization, frequency and positional degrees of freedom~\cite{Lanyon09}, and most superconducting quantum circuits~\cite{Devoret13}. The transmon~\cite{Koch07}, phase~\cite{Martinis02} and capacitively-shunted flux~\cite{Steffen10} qubits are weakly anharmonic oscillators with logical transition frequency $\wq$ and nearest leakage transition frequency $\wl$ detuned by $|\Delta|=|\wl-\wq| \sim 0.1 \times \wq$.  In these superconducting systems, temporarily occupying  levels outside the computational subspace offers the key to fast and efficient multi-qubit operations such as conditional-phase~\cite{Strauch03, DiCarlo09} and Toffoli gates~\cite{Fedorov11, Reed12}, and high-fidelity single-shot readout~\cite{Mallet09}.

The benefits of using multi-level structures for quantum computing are balanced by more challenging single-qubit control. When driving an individual effective qubit with a resonant pulse at $\wq$, the anharmonicity $|\Delta|$ imposes a practical limit on the maximum speed of gate operations, marking the transition from decoherence- to leakage-dominated errors. While theoretical optimal control has broken the speed limit using non-analytic pulses~\cite{Motzoi09}, analytic pulses with few tuning parameters are preferred by experimentalists for  ease of implementation and tuning. Keeping leakage-induced errors below the $1\%$ fault-tolerance threshold of modern error-correcting schemes~\cite{Fowler12} imposes the necessary but insufficient condition $\tg \gtrsim 2\pi/|\Delta|$ on the single-qubit gate time $\tg$. Interestingly, the standard Gaussian envelope is insufficient despite satisfying the minimal time-frequency uncertainty product. Proposed~\cite{Motzoi09, Gambetta11, Motzoi13} DRAG (Derivative Removal by Adiabatic Gate) pulses combining Gaussian and derivative-of-Gaussian envelopes on the in- and out-of-phase quadratures have been widely adopted following validation with phase~\cite{Lucero10} and transmon~\cite{Chow10b} qubits in one- and two-qubit devices. To date, the combination of DRAG and improved coherence has achieved average single-qubit gate errors of $0.08\%$ in transmon qubits~\cite{Barends14a}.

Moving forward, it is imperative to preserve high-quality single-qubit control as more effective qubits are crowded in a fixed frequency range. In architectures such as  2D~\cite{Wallraff04}  and 3D~\cite{Paik11} circuit QED which exploit a common feedline or coupled resonator to drive multiple  qubits, control drives couple almost equally to addressed and unadressed qubits. In this regime of near-unity cross-talk,  the absolute detuning $|\delta|$ between the logical transition of one qubit and the leakage transition of its frequency neighbor sets an even lower speed limit when $|\delta|<|\Delta|$. In order to ease coherence time requirements, it is therefore important to design analytic pulses with $\tg \sim 2\pi/|\delta|$ which avoid leakage in both the addressed qubit and its neighbor (henceforth termed internal and external leakage). To this end, Schutjens \emph{et al.}~\cite{Schutjens13} have recently developed Wah-Wah control (Weak AnHarmonicity With Average Hamiltonian), combining DRAG with sideband modulation in a four-parameter pulse.

In this article, we present the experimental validation of leakage-avoiding Wah-Wah control at the speed limit of a multi-transmon 2D circuit QED processor. We create a bias condition with $\delta/2\pi= 57~\MHz$ and demonstrate avoidance of both external and internal leakage at gate times $16~\ns\leq\tg \leq 24~\ns$. Stroboscopic population measurements show that DRAG-only pulsing induces significant net population in the third level of the unaddressed transmon, while Wah-Wah ensures all population returns to the computational subspace by the end of the pulse. Using a variant of standard randomized benchmarking~\cite{Knill08}, we show alternating individual control of both qubits with $0.8-1.0\%$ average error per computational step. Finally, we use quantum process tomography to demonstrate decoherence-limited idling of the unaddressed qubit as the other undergoes a Wah-Wah pulse train. Optimization of the four-parameter, analytic Wah-Wah pulse shape is straightforward and accelerated by a simple model of the system Hamiltonian using independently measured parameters. Our results establish Wah-Wah control as an important tool for scalability, allowing control of frequency-crowded effective qubits at threshold without dedicated control lines.

\begin{figure}[t!]
\includegraphics{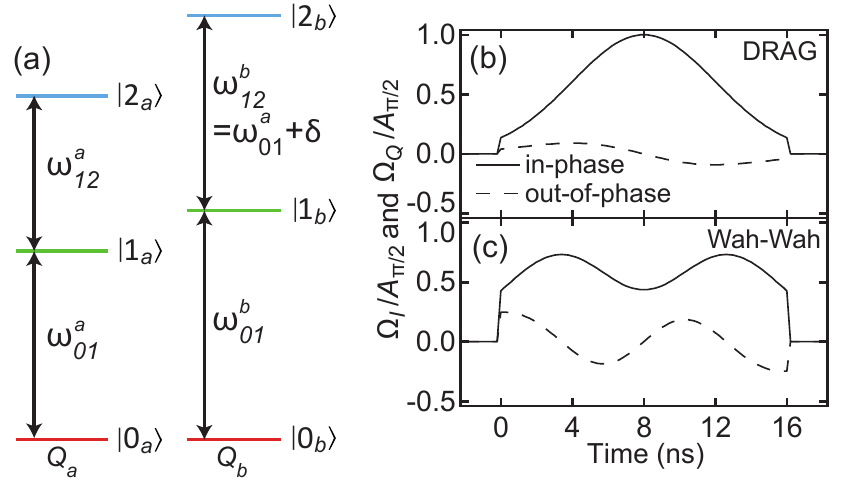}
\caption{(color online).  (a) Energy level diagram for transmons $\QA$ and $\QB$ (not to scale). The ground $\ket{0_k}$  and first-excited $\ket{1_k}$ states of $Q_k$ define a qubit subspace. The $\QA$ qubit transition frequency and the $\QB$ leakage transition frequency differ by $\delta=\wtB-\wA=2\pi \times 57~\MHz$. (b-c) Comparison of in- and out-of-phase quadrature envelopes $\Omega_I$ and $\Omega_Q$, respectively, for optimized DRAG and Wah-Wah $\pi/2$ pulses on $\QA$ (gate time $\tg=16~\ns$). For DRAG (b), $\Omega_I$ and $\Omega_Q$ are Gaussian and derivative-of-Gaussian, respectively [Eqs.~(1)-(2) with  $\sigma=4~\ns$, $\Am=0$, $\beta=0.6~\ns$]. For Wah-Wah (c), $\Am=0.9$, $\wm/2\pi=25~\MHz$, $\beta=1.85~\ns$.
}
\end{figure}

We focus on two transmons ($\QA$ and $\QB$) within a four-transmon, five-resonator 2D cQED processor of similar design to that of Ref.~\onlinecite{Saira14}. Resonant control and readout pulses for all qubits are applied via one feedline coupling to readout resonators connecting  to one qubit each. Using local flux control, we bias $\QA$ and $\QB$ to logical transitions  $\left(\wA,\wB\right)/2\pi=\left(6.347, 6.750\right)~\GHz$, and corresponding leakage transitions $\left(\wtA,\wtB\right)/2\pi=\left(5.980, 6.404\right)~\GHz$, making $\delta=\wtB-\wA=2\pi \times 57~\MHz$ [Fig.~1(a)]. Thus, we expect~\cite{Schutjens13} DRAG pulses [Fig.~1(b)] targeting $\QA$ to induce significant leakage from $\ketB{1}$ to $\ketB{2}$ for $\tg \lesssim 2\pi/\delta \sim 20~\ns$. Indeed, we note that just four back-to-back $\QA$ DRAG $\pi$ pulses already leak $\sim50\%$ of the initial population in $\ketB{1}$ to $\ketB{2}$ for $\tg=16~\ns$ (Fig.~2). Note that to within the few-percent accuracy limited by state preparation and measurement errors (SPAM), DRAG pulses do successfully avoid internal leakage in $\QA$, as expected~\cite{Motzoi09}.

Using similar measurements, we now  attempt to also avoid external leakage using the additional sideband modulation characteristic of Wah-Wah pulse envelopes [Fig.~1(c)]~\cite{Schutjens13}:
\begin{eqnarray}
\label{eq:wahx}
&&\Omega_I(t)=A_\theta e^{-(t-\frac{\tg}{2})^2/(2\sigma^2)}\left[ 1-\Am\cos \left(\wm\left(t-\frac{\tg}{2}\right)\right)\right]\nonumber\\
\\
\label{eq:wahy}
&&\Omega_Q(t)=\beta \dot{\Omega}_I(t).
\end{eqnarray}
Here, $\Omega_I$ and $\Omega_Q$ are the pulse envelopes in the in- and out-of-phase quadratures, and amplitude $A_\theta$ determines the rotation angle $\theta$. The theory predicts that sideband modulation of the conventional Gaussian envelope in $\Omega_I$ can mitigate external leakage for suitably chosen modulation amplitude $\Am$ and frequency $\wm$. Just as in DRAG, keeping $\Omega_Q$  proportional to the time derivative of $\Omega_I$ should prevent internal leakage in $\QA$ upon optimizing the scaling parameter $\beta$. Figure 2 provides the experimental confirmation of external and internal leakage mitigation to within the accuracy allowed by SPAM. $\QA$ Wah-Wah $\pi$ pulses with manually-optimized $\Am$, $\wm$, $\Api$, and $\beta$ populate $\ketB{2}$ only temporarily, returning all population to $\ketB{1}$ by the end of each pulse. (Details of the Wah-Wah pulse tune-up procedure are provided in the Supplement~\cite{SOM}). A numerical simulation of the system dynamics, which truncates the Hamiltonian at three levels per transmon, shows good correspondence with the manually optimized pulse parameters. To test its utility, we used the simulation to obtain first estimates of pulse parameters at two other $\QB$ bias points with even tighter separation of their logical frequencies ($\delta/2\pi=-60~\MHz$ and $-80~\MHz$). We also obtained similar internal and external leakage mitigation for $\tg=16~\ns$ (data not shown).
\begin{figure}
\includegraphics{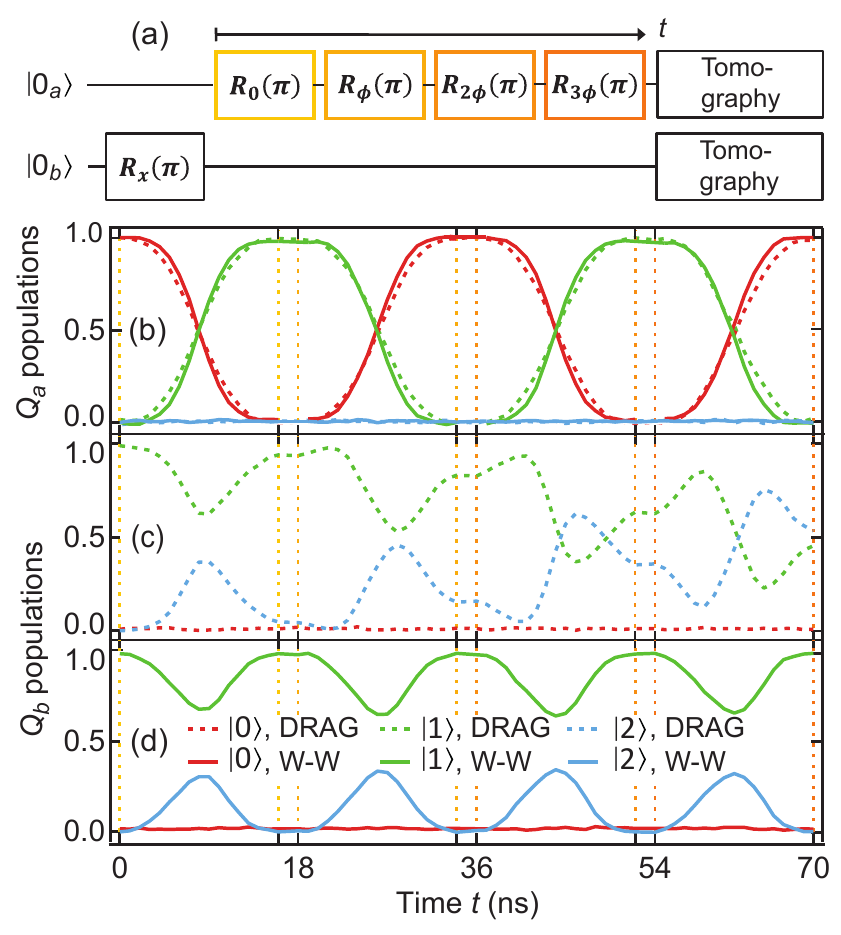}
\caption{(color online). Measured evolution of level populations in $\QA$ and $\QB$ during four consecutive $\QA$ $\pi$ pulses with either DRAG and Wah-Wah envelopes ($\tg=16~\ns$).  (a) Pulse sequence. Pulses are separated by a $\tb=2~\ns$ buffer. Level populations at time $t$ are obtained by truncating any ongoing $\QA$ pulse and commencing tomographic operations after a buffer time $\tb$. (b) Evolution of $\QA$ levels. Neither DRAG (dashed curves) nor Wah-Wah (solid curves) pulses drive the leakage transition in $\QA$. (c-d) Evolution of $\QB$ levels during DRAG (c) and Wah-Wah (d) $\QA$ pulses. DRAG pulsing drives the $\QB$ leakage transition. The chosen relative phase $\phi=237^\circ$ between subsequent $\pi$ pulses exacerbates the net leakage. In contrast, Wah-Wah pulses populate $\ketB{2}$ temporarily, returning the population to $\ketB{1}$ by the end of each pulse.
}
\end{figure}

\begin{figure}[t!]
\includegraphics{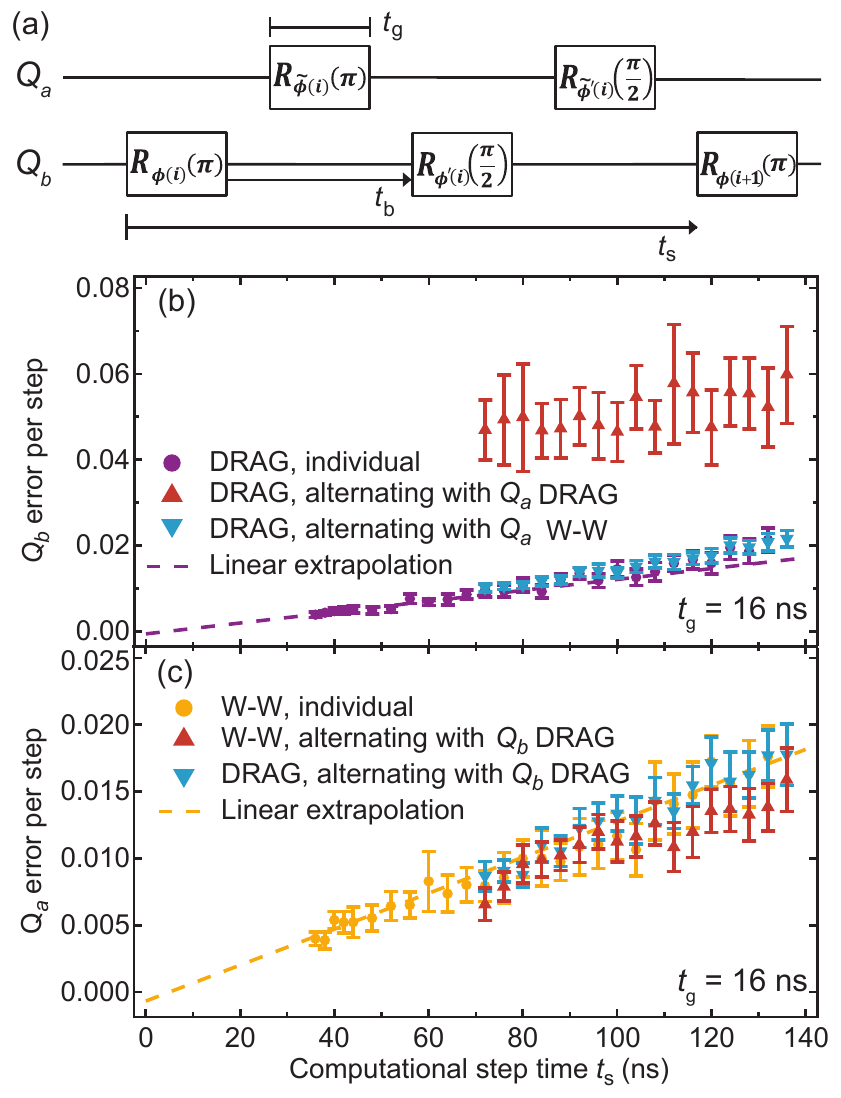}
\caption{(color online).
Demonstration of decoherence-limited single-qubit control via individual and alternating randomized benchmarking.
(a) Pseudo-random control pulses are applied to either one transmon or to both in alternating fashion.  The average error per computational step ($\EPG$) is extracted as a function of the computational step time $\ts=2\tg+2\tb$ (without and with a symmetrically timed RB pulse on the other transmon during buffer). (b) $\EPG$ for optimized $\QB$ DRAG pulses. The linear $\ts$ dependence observed without $\QA$ pulses extrapolates to $\left(-6 \pm 9\right)\times 10^{-4}$ at $\ts=0$, indicating decoherence-limited control. The fits are done to the first ten data points. The alternating $\EPG$ matches the individual $\EPG$ for $\QA$ Wah-Wah. In contrast, alternating with $\QA$ DRAG pulsing worsens the $\EPG$ to $\sim5~\%$. The large error bars reflect  high sensitivity of $\ketB{2}$ leakage to the particular sequence of $\QA$ rotations within the randomization. (c) $\EPG$ for optimized $\QA$ DRAG and Wah-Wah pulses. Overlapping results are obtained without and with alternating RB DRAG pulses on $\QB$. The observed linear $\ts$ dependence of $\EPG$ extrapolates to $\left(-7 \pm 9\right)\times 10^{-4}$ at $\ts=0$, indicating that $\QA$ control is also decoherence limited.
}
\end{figure}

While Wah-Wah pulsing on $\QA$ successfully mitigates net leakage in $\QB$, the temporary excursion of quantum amplitude from $\ketB{1}$ to $\ketB{2}$ induces a relative phase between levels $\ketB{0}$ and $\ketB{1}$, i.e., a $z$ rotation in the $\QB$ qubit subspace. This induced phase is a deterministic function of the $\QA$ pulse parameters defined above. Thus, we can compensate it already in pulse synthesis by adjusting the phase of all subsequent $\QB$ pulses, in the style of virtual $z$ gates~\cite{Steffen00}. To calibrate the phase shift, we embed several consecutive Wah-Wah $\QA$ pulses (either all $\pi$ or $\pi/2$) into the second wait period in a standard echo sequence on $\QB$ ($\pi/2$, wait, $\pi$, wait, $\pi/2$). The final $\pi/2$ rotation translates the acquired phase into a population difference between $\ketB{0}$ and $\ketB{1}$. We observe that the induced $\QB$ phase is independent of rotation axis and linear in the number of $\QA$ pulses of a given type: $8.2^\circ$ ($38^\circ$) per $\pi/2$ ($\pi$) pulse.  We perform a similar calibration of compensating $z$ gates on $\QA$ for DRAG pulses applied to $\QB$. Even though $\left(\wtA-\wB\right)/2\pi=-770~\MHz$ and $\QB$ DRAG pulsing does not produce $\QA$ leakage (shown below), there is phase accrual in the $\QA$ qubit subspace: $2.5^\circ$ ($9.3^\circ$) per $\pi/2$ ($\pi$) pulse.

In order to test both leakage mitigation and  phase compensation to higher accuracy than allowed by SPAM, we employ randomized benchmarking (RB). The single-qubit protocol first proposed and implemented by Knill \emph{et al.}~\cite{Knill08} (standard RB) provides a valuable baseline for gate errors on the addressed qubit, without concern for the unaddressed one. In standard RB~\cite{Knill08, Chow09}, one applies pseudo-random sequences of consecutive $\pi$ and $\pi/2$ pulses to $\QK$ and measures the decay of fidelity to the ideal final state of that qubit (always $\ketK{0}$ or $\ketK{1}$) as the number of pulses is increased [Fig.~3(a)]. This decay allows extracting~\cite{Magesan12,SOM} the average error per computational step ($\EPG$), where computational step is defined as a pair of back-to-back $\pi$ and $\pi/2$ pulses~\cite{Knill08}.  We extract $\EPG$ as a function of the step time  $\ts=2(\tg+\tb)$ by varying the buffer time $\tb$ between pulses ($\tb\geq 2~\ns$). To within statistical error, a linear fit of $\EPG(\ts)$ for $\QA$ ($\QB$) at short $\ts$  extrapolates to the origin [Figs.~3(b)-(c)]. This observation suggests that the minimal $0.4\%$ ($0.4\%$) $\EPG$ is already decoherence limited.

\begin{figure}[t!]
\includegraphics{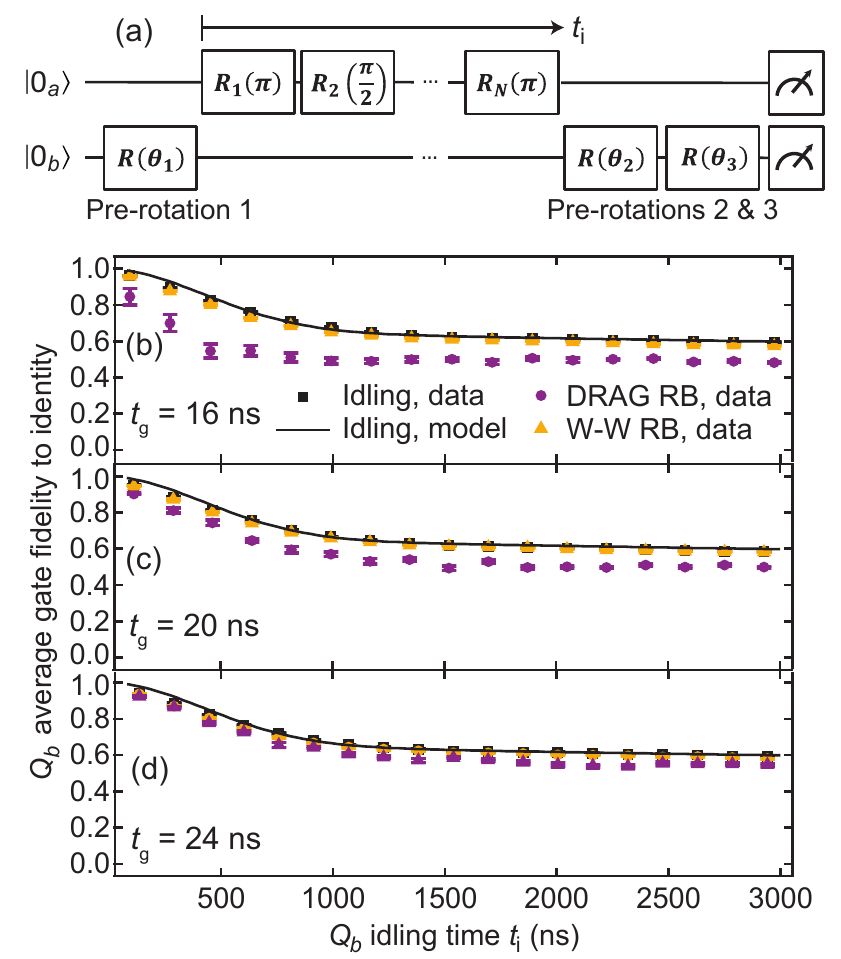}
\caption{(color online). $\QB$ average gate fidelity to the identity operation during various types of idling: no $\QA$ pulsing, $\QA$ DRAG RB, and $\QA$ Wah-Wah RB.  (a) Pulse sequence. Initial and final rotations on $\QB$ are used for QPT of the evolution of $\QB$ qubit state to the three-level subspace. Results are shown for gate times (b) $\tg=16~\ns$, (c) $20~\ns$, and (d) $24~\ns$ ($\tb=2~\ns$ fixed). Model curves for true idling take into account the measured $\QB$ qubit relaxation and dephasing times~\cite{SOM}. The model assumes dephasing dominated by $1/f$ noise~\cite{Yoshihara06}.
The $\QB$ average gate fidelity to identity during $\QA$ Wah-Wah RB (circles) is indistinguishable from that of true idling. In  contrast, $\QA$ DRAG RB (triangles) deteriorates fidelity at shorter $\ti$.
}
\end{figure}

With the $\EPG$ baselines from standard RB in place, we now employ alternating RB to investigate whether control of either qubit can remain decoherence limited when pulses are interleaved on the other. We apply an RB pulse on one qubit during the buffer ($\tb\geq \tg+4~\ns$) for the other, and perform virtual $z$-gate compensation for all pulses. Alternating RB performed with DRAG pulses on both $\QA$ and $\QB$ has no impact on $\QA$, but increases the $\QB$ $\EPG$ to $\sim5\%$ [Fig.~3(b)]. However, by using Wah-Wah pulses on $\QA$, we recover the decoherence limited baselines simultaneously on both qubits. Similar results for $\tg=20~\ns$ and $24~\ns$ are presented in the Supplement~\cite{SOM}.

As the final test of whether Wah-Wah pulsing on $\QA$ affects $\QB$, we perform quantum process tomography (QPT) of $\QB$ under various idling conditions: no pulses on $\QA$ (true idling), $\QA$ DRAG RB, and $\QA$ Wah-Wah RB (Fig.~4). The process map is a $9\times 4$ transfer matrix~\cite{Chow12} relating an initial reduced density matrix in the $\QB$ qubit subspace to a final reduced density matrix in the three-level~\cite{Bianchetti10} subspace. The $\QB$ average gate fidelity to identity for true idling decreases consistently with an analytic model~\cite{SOM} based on measured $\QB$ relaxation and dephasing rates. As expected, $\QB$ idling is compromised under $\QA$ DRAG RB, and worsens the shorter $\tg$. An analysis discerning contributions from population transfer and dephasing errors~\cite{SOM} confirms that the loss of idling fidelity is limited by induced leakage and not by imperfect $z$-gate compensation. Remarkably, the idling fidelity under $\QA$ Wah-Wah RB is nearly identical to the true idling fidelity for all $\tg$, further demonstrating the usefulness of the new control method.

In summary, we have shown experimentally, using level-population measurements, RB, and QPT, that Wah-Wah control~\cite{Schutjens13} successfully mitigates crosstalk-induced leakage in a crowded spectrum of transmon qubits at gate times where the widely-adopted DRAG control fails. Wah-Wah control therefore represents a step towards scalability in multi-qubit architectures by allowing selective control of an increasing number of effective qubits without an equal addition of control lines.  Wah-Wah builds sideband modulation to DRAG without sacrificing the analytic nature of the two-quadrature pulse shape, adding only two easily optimized parameters.  Our demonstration in circuit QED takes place in the worst-case regime of unity crosstalk, with control pulses applied via one common feedline coupling equally to the addressed qubit and the unaddressed nearest-frequency neighbor.  Looking forwards, as two-qubit gate and measurement times~\cite{Sank14} decrease to that of single-qubit gates, the current restriction to non-overlapping control of frequency-neighboring qubits could ultimately bottleneck the clock cycle in surface-code quantum error correction~\cite{Fowler12}.  A useful generalization of Wah-Wah, however, would allow simultaneous control of two qubits at small $\left|\delta\right|$.

\begin{acknowledgments}
We thank D.~J.~Egger, R.~Schutjens and F.~K.~Wilhelm for many helpful discussions, N.~K.~Langford and D.~Rist\`{e} for helpful comments on the manuscript, and D.~J.~Thoen and T.~M.~Klapwijk for NbTiN thin films. This work is funded by EU FP7 project SCALEQIT, the Netherlands Organization for Scientific Research (NWO, VIDI scheme), and a Marie Curie Career Integration Grant. V.V. was supported by the Research Foundation of Helsinki University of Technology.
\end{acknowledgments}

\bibliographystyle{prl}
\bibliography{../../../TeX/References/References_cQED}

\end{document}


\title{Supplement to ``Mitigating information leakage in a crowded spectrum of weakly anharmonic qubits''}
\author{V.~Vesterinen}
\affiliation{VTT Technical Research Centre of Finland, P.O. Box 1000, 02044 VTT, Finland}
\affiliation{Kavli Institute of Nanoscience, Delft University of Technology, P.O. Box 5046,
2600 GA Delft, The Netherlands}
\author{O.-P.~Saira}
\affiliation{Kavli Institute of Nanoscience, Delft University of Technology, P.O. Box 5046,
2600 GA Delft, The Netherlands}
\author{A.~Bruno}
\affiliation{Kavli Institute of Nanoscience, Delft University of Technology, P.O. Box 5046,
2600 GA Delft, The Netherlands}
\author{L.~DiCarlo}
\affiliation{Kavli Institute of Nanoscience, Delft University of Technology, P.O. Box 5046,
2600 GA Delft, The Netherlands}
\date{\today}

\maketitle

\section{Experimental details}

\subsection{Device}  
\begin{figure*}[t!]
\includegraphics{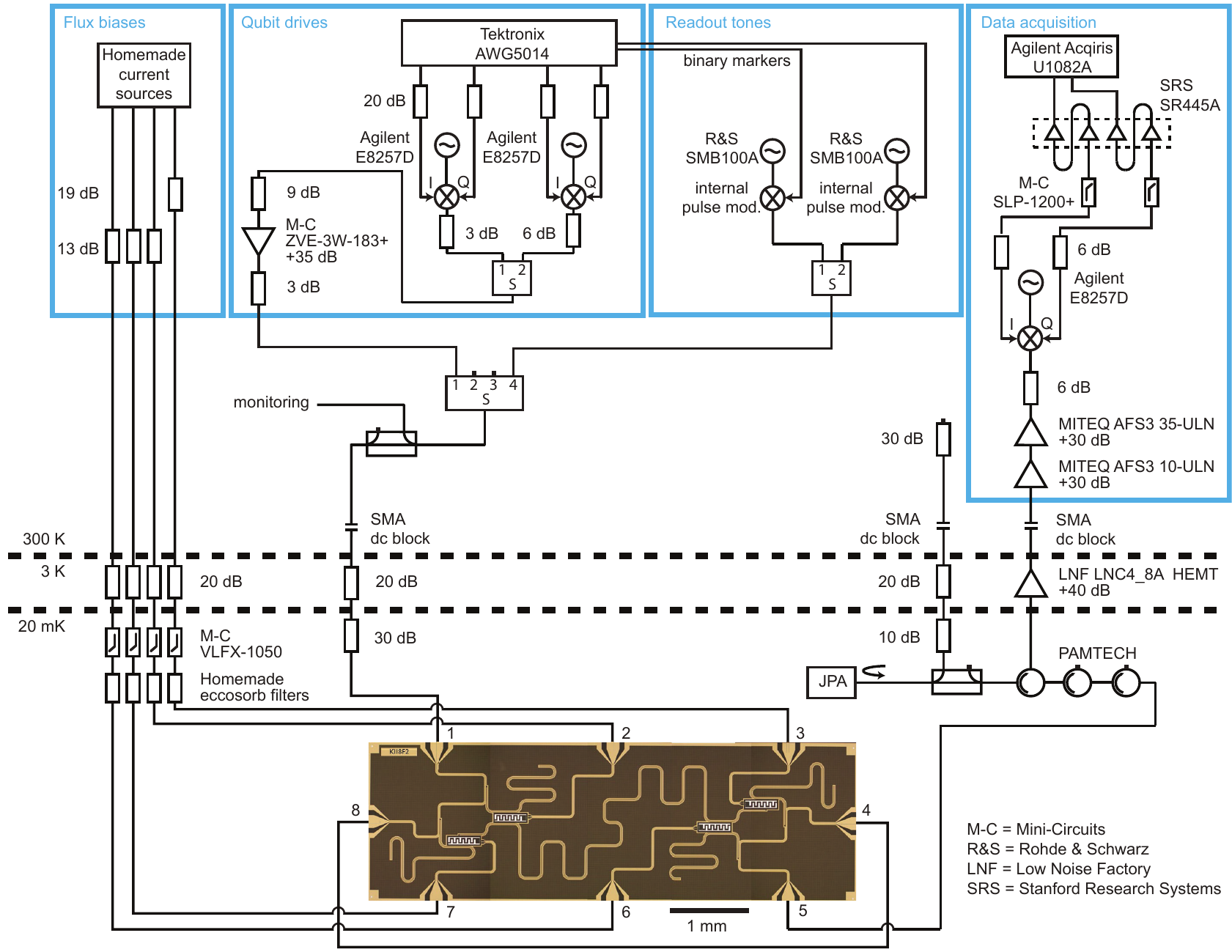}
\caption{Device and experimental setup. The $2~\mm \times 7~\mm$ chip is cooled to $20~\mK$ in a $\Hethree/\Hefour$ dilution refrigerator (Leiden Cryogenics CF-450). The chip ports are labeled 1 through 8. Low-pass filtered d.c. currents generating static flux biases for the transmons enter through ports $2$ (inactive transmon, transition frequency $4.31~\GHz$), $3$ ($\QA$), $6$ (also inactive, transition frequency $7.25~\GHz$), and $7$ ($\QB$). All microwave control and readout pulses are applied at the single-feedline input port (1). Feedline ports $8$ and $4$ are externally connected by a short coaxial cable.
Transmon readout is performed applying two simultaneous square-envelope pulses ($1~\us$ duration) near the fundamental frequencies of the dedicated resonators coupled to $\QA$ and $\QB$. The transmitted feedline signal exiting at port 5 is routed by circulators (Pamtech) past a Josephson parametric amplifier (JPA, unpumped and unused) and into a HEMT amplifier (Low Noise Factory) at $3~\K$.  Two room-temperature amplifiers (Miteq) further amplify the readout signals, which are subsequently demodulated with an IQ mixer (Marki Microwave). The mixer local oscillator frequency is chosen equal to the measurement frequency of one of the two readout resonators (thus, $0~\Hz$ IF) and is $100~\MHz$ offset from the second. The IF signals are amplified (Stanford Research Systems), digitized (Agilent), and homodyne detected digitally to complete the simultaneous readout of both transmons. DRAG and Wah-Wah pulse envelopes for resonant qubit control are generated by a Tektronix AWG5014 arbitrary waveform generator. We employ $\pm50~\MHz$ single-sideband modulation to prevent spurious transmon driving by leakage of the local oscillator in the IQ mixers used for pulse up-conversion.
}
\label{fig:circuit}
\end{figure*}

The chip is a four-transmon, five-resonator 2D cQED quantum processor of nearly identical design and fabrication as that presented in Ref.~\onlinecite{Saira14}. An optical image of the device and detailed schematic of the setup are shown in Fig.~\ref{fig:circuit}. A high-$Q$ resonator bus ($5.16~\GHz$ fundamental) couples to every transmon, while dedicated resonators, each dispersively coupled to one transmon, allow individual readouts via a common feedline. Transmon transition frequencies are individually controlled by dedicated flux-bias lines, each short-circuited near one transmon SQUID loop. Throughout this experiment, $\QA$ was biased at its flux-insensitive point, where  $\wA/2\pi=6.347~\GHz$ and $\DA/2\pi=-357~\MHz$. The $\QA$ readout resonator has a fundamental frequency of $\wrA/2\pi=7.7042~\GHz$ (for $\QA$ in $\ketA{0}$), a coupling-limited linewidth of $\kappa_a/2\pi=1.5~\MHz$, and a dispersive coupling strength of $\chi_a/\pi=-1.3~\MHz$. Three bias points were explored for $\QB$ (Table \ref{table:biaspoints}). The other two (inactive) transmons on the chip were biased at $4.31~\GHz$ and $7.25~\GHz$ throughout.

\begin{table}[h!]
\caption{Summary of $\QB$-related device parameters at the three bias points explored.}
\begin{tabular}{l|c|c|c|}
Bias point                 & 1 & 2 & 3\\
\hline
$\wB/2\pi$ (GHz)         & $6.750$ & $6.636$ & $6.616$\\
$\DB/2\pi$ (MHz)         & $-346$ & $-349$ & $-350$\\
$\delta/2\pi$ (MHz)      & $57$ & $-60$ & $-81$\\
$\wrB/2\pi$ (GHz)& 7.8181 &  7.8178  & 7.8177\\
$\kappa_b/2\pi$ (MHz)& 1.8& N.A. & N.A. \\
$\chi_b/\pi$ (MHz)      & -1.7& N.A. & N.A. \\
\end{tabular}
\label{table:biaspoints}
\end{table}

\subsection{Additional device parameters at bias point 1}
For bias point 1, where all shown data were taken, we calibrated several device parameters needed as input for simulation (discussed below). Measured relaxation and dephasing times of $\QA$ and $\QB$ are listed in Table~\ref{table:coherence}. The Rabi frequencies of equal-amplitude drives resonant with the $\ket{0}\leftrightarrow\ket{1}$ and $\ket{1}\leftrightarrow\ket{2}$ transitions of each transmon are listed in Table~\ref{table:lambdas},
normalized to that of the $\ketA{0}\leftrightarrow \ketA{1}$ transition. From these, we estimate the relative coupling strength of a drive centered at $\wA$ to the four transitions by simulating the filter functions of the two readout resonators at $\wA$ using Microwave Office.

\begin{table}
\centering
\caption{
Measured qubit relaxation  $T_{1(1\rightarrow0)}$, Ramsey $T_2^{\mathrm{Ramsey}}$, and echo $T_2^{\mathrm{echo}}$ times at bias point 1. The measured relaxation time $T_{1(2\rightarrow1)}$ from second to first excited state in each transmon is also listed.}
\begin{tabular}{l|c|c|}
Transmon & $\QA$ & $\QB$\\
\hline
$T_{1(1\rightarrow0)}~(\us)$    & $7.65$ & $5.65$\\
$T_{1(2\rightarrow1)}~(\us)$    & $4.18$ & $3.66$\\
$T_2^{\mathrm{Ramsey}}~(\us)$   & $2.13$ & $0.64$\\
$T_2^{\mathrm{echo}}~(\us)$    & $2.33$ & $1.40$\\
\end{tabular}
\label{table:coherence}
\end{table}

\begin{table}
\centering
\caption{Measured Rabi frequencies for equal-amplitude drives resonant with the $\ketK{0}\leftrightarrow\ketK{1}$ and $\ketK{1}\leftrightarrow\ketK{2}$ transitions, and estimated coupling strength of the $\wA$ drive to the four transitions. All values are normalized to that of the $\ketA{0}\leftrightarrow \ketA{1}$ transition.
}
\begin{tabular}{l|c|c|}
Transmon $\QK$& $\QA$ & $\QB$\\
\hline
Rabi frequency at $\wK$   & $1$ & $0.90$\\
Rabi frequency at $\wtK$ & $1.42$ & $1.25$\\
Coupling $\lambda_1^k$ of $\wA$ drive to $\ketK{0}\leftrightarrow\ketK{1}$ & 1 & 0.5\\
Coupling $\lambda_2^k$ of $\wA$ drive to $\ketK{1}\leftrightarrow\ketK{2}$ & 2.4 & 1.2
\end{tabular}
\label{table:lambdas}
\end{table}

\subsection{Transmon readout}

\emph{Multiplexed readout.} Simultaneous, independent readouts of $\QA$ and $\QB$ were performed by applying square-envelope tones ($1~\us$ duration) at $\wmA=\wrA+2\chi_a$ and $\wmB\approx\wrB+2\chi_b$ to the feedline, respectively. To preserve the phase of measurement tones between experiment repetitions, we ensured that $\omega_m^\Delta = \omega_m^b - \omega_m^a$ was an integer multiple of $2\pi/t_\mathrm{rep}$, where $t_\mathrm{rep}=100~\us$ is the experiment repetition time. The amplified feedline output was demodulated by an IQ mixer, low-pass filtered (corner frequency $1.2~\GHz$) and digitized at $\delta t = 1~\ns$ sampling interval (see Fig.~\ref{fig:circuit} for the complete readout chain). The mixer local oscillator frequency was chosen equal to $\omega_m^a$. The two quadratures for $\QA$ readout were obtained by filtering the $I[n]$ and $Q[n]$ streams with
an averager rejecting all multiples of $\omega_m^\Delta$. For $\QB$ readout, the two quadratures were derived from  $I[n]$ by computing $I[n] \cos (\omega_m^\Delta\, n \delta t)$ and $I[n] \sin (\omega_m^\Delta\, n \delta t)$ and filtering with the averager. Finally, these four signals were integrated over the $1~\us$ interval to obtain the four voltages $V_I^a$, $V_Q^a$, $V_I^b$, and $V_Q^b$. The single-shot readout fidelities of $\QA$ and $\QB$ were $63\%$ and $65\%$, respectively. We note that the parametric amplifier present in the readout chain was not employed as we did not require high readout fidelity for this experiment.

\emph{Measurement model.} We relate the average integrated voltages $\mVIK$ and $\mVQK$ to the level populations of transmon $\QK$ using the model~\cite{Filipp09}
\[
\langle V_i^k \rangle = \tr (\rho_k M_i^k),
\]
where
\begin{equation}
M_i^k = \beta_{i0}^k \Pi_0^k + \beta_{i1}^k \Pi_1^k + \beta_{i2}^k \Pi_2^k,
\end{equation}
$\hat{\Pi}_l^{k}=\ket{l_{k}}\bra{l_{k}}$, and $\rho_k$ the reduced qutrit density matrix of $\QK$. We calibrate the coefficients  $\beta_{i,j}^k$ by measuring $\langle V_i^k\rangle$ immediately after preparing $\ketK{0}$, $\ketK{1}$, and $\ketK{2}$. We prepare the last two states using optimized DRAG pulses $\Rk{x,01}{\pi}$ and $\Rk{x,12}{\pi} \Rk{x,01}{\pi}$, respectively (see below for pulse details).
Here, subscripts $01$ and $12$ indicate rotations in the $\{\ketK{0},\ketK{1}\}$ and $\{\ketK{1},\ketK{2}\}$ subspaces, respectively.

\emph{Extraction of level populations.}
To extract the level populations $P_j^k=\tr (\rho_k \Pi_j^k)$ in transmon $\QK$, we solve the set of three linear equations given by  $\mVIK$, $\mVQK$ and the (assumed) constraint $\sum_{i=0}^2 P_j^k=1$.
For the stroboscopic measurements in Figs.~2 and S2, we enhance the accuracy of $\QB$ level population measurements by increasing the number of linear equations to 7. We measure $\mVIB$ and $\mVQB$ with measurement pre-rotations $I$, $\Rb{x,01}{\pi}$ and $\Rb{x,12}{\pi}$, and perform unweighted least-squares inversion.  $\QA$ level population measurements in Fig.~2(b) are performed with $\QB$ pre-rotation $I$.

\subsection{Pulse tuning}

Tune-up of $\QA$ and $\QB$ pulses at each bias point and $\tg$ began with a manual optimization of DRAG pulses on each transmon without concern for their effect on the other. To facilitate the fine tuning of the pulse amplitude $A_{\theta}$ ($A_{\pi/2}=A_\pi/2$) and the DRAG parameter $\beta$, we used a test sequence similar to those in Refs.~\onlinecite{Chow10} and \onlinecite{ReedPhD13}. Each segment of the sequence applies one of 21 different pairs of pulses (each drawn from $\{I,\Rk{x,01}{\pi/2},\Rk{y,01}{\pi/2}, \Rk{x,01}{\pi}, \Rk{y,01}{\pi}\}$) to $\ketK{0}$ and immediately performs measurement on $\QK$. The pulse pairs are picked and ordered so that the $\QK$ qubit Bloch vector  is ideally left at the north pole, equatorial plane or south pole of the Bloch sphere, in progression.  The deviations from a three-level staircase in $\langle V_i^k \rangle$ provide a useful footprint of tune-up errors.

When turning on Wah-Wah pulsing of $\QA$ at bias point 1, we followed a manual optimization procedure. For each choice of sideband-modulation parameters $\Am$ and $\wm$, we first estimated the amplitude needed to preserve the
area under the in-phase quadrature envelope. We multiplied the DRAG $A_{\theta}$ by  $I_\mathrm{DRAG}/(I_\mathrm{DRAG}-\Am I_\mathrm{W-W})$, where
\begin{eqnarray}
&&I_\mathrm{DRAG}=\int_{0}^{2\sigma} e^{-t^2/2\sigma^2}dt,\\
&&I_\mathrm{W-W}=\int_{0}^{2\sigma} e^{-t^2/2\sigma^2}\cos (\wm t)dt.
\end{eqnarray}
We found this method to be accurate to $\pm 0.4\%$. Using the test sequence above (but implemented with Wah-Wah pulses), we next tuned $\beta$. Finally, similarly to Figs.~2 and ~\ref{fig:strobopi2}, we measured the $\Ptwob$ produced by a sequence of four back-to-back $\QA$ Wah-Wah pulses. In this way, we manually found the Wah-Wah parameters ($A_{\theta}$, $\beta$, $\Am$, and $\wm$) pulses that produce minimal $\Ptwob$ while performing the intended $\QA$ operation. The procedure was separately performed for $\theta=\pi$ and $\pi/2$ pulses.  Once a good correspondence was established between manually optimized pulse parameters and those suggested by simulation, we increasingly relied on simulation to fix $\Am$ and $\wm$, and only manually tuned $\beta$.  A summary of optimal $\QA$ DRAG, $\QA$ Wah-Wah, and $\QB$ DRAG pulse parameters at the three bias points and several $\tg$ is provided in Table~\ref{table:deltas}.

We finally note that to calibrate DRAG pulses in the $\{\ketK{1},\ketK{2}\}$ subspace, we modified the test sequence to apply a pre-optimized $\Rk{x,01}{\pi}$ before applying the pulse pairs (each drawn from $\{I,\Rk{x,12}{\pi/2},\Rk{y,12}{\pi/2}, \Rk{x,12}{\pi}, \Rk{y,12}{\pi}\}$). No additional mixers were required for pulse up-conversion as both $\wK$ and $\wtK$ could be reached by single-sideband modulation.

\begin{table}[h!]
\caption{Parameters of optimized $\QA$  and $\QB$ pulses at three bias points.}
\begin{tabular}{l|l|l|c|c|c|}
\multicolumn{3}{c}{}&\multicolumn{3}{|c|}{Bias point}\\
\cline{4-6}
\multicolumn{3}{c|}{}& 1 & 2 & 3\\
\hline
&&$\Am$& $0.9$ & $0.3585$ &  $-0.8$\\
&W-W$\Ra{\hat{n},01}{\pi}$&$\wm/2\pi$ (MHz) &$12.5$ & $99.6$ & $17.5$ \\
&&$\beta$ (ns) & $0.9$ & $0.2$ & $0.2$\\
\cline{2-6}
&&$\Am$&$0.9$ &$0.6743$ &$0.5$ \\
$\tg=16~\ns$&W-W $\Ra{\hat{n},01}{\pi/2}$&$\wm/2\pi$ (MHz) &$25$& $76.5$  &$25$ \\
&&$\beta$ (ns)& $1.85$ & $0.2$ & $0.3$\\
\cline{2-6}
&DRAG $\Ra{\hat{n},01}{\theta}$&$\beta$ (ns)& $0.6$ & $0.2$ & $0.2$\\
\cline{2-6}
&DRAG $\Rb{\hat{n},01}{\theta}$&$\beta$ (ns)& $0.7$ & $-0.1$ & $-0.1$\\
\hline
&&$\Am$ & $0.23$ &&\\
&W-W$\Ra{\hat{n},01}{\pi}$&$\wm/2\pi$ (MHz)&$13.8$  &&\\
&&$\beta$ (ns)& $0.62$ &  & \\
\cline{2-6}
&&$\Am$ &$0.68$  && \\
$\tg=20~\ns$&W-W$\Ra{\hat{n},01}{\pi/2}$&$\wm/2\pi$ (MHz)  & $23.8$ &&\\
&&$\beta$ (ns)& $0.95$ & & \\
\cline{2-6}
&DRAG $\Ra{\hat{n},01}{\theta}$&$\beta$ (ns)& $0.6$ & & \\
\cline{2-6}
&DRAG $\Rb{\hat{n},01}{\theta}$&$\beta$ (ns)& $0.68$ & & \\
\hline
&&$\Am$ & $-0.65$ &&\\
&W-W$\Ra{\hat{n},01}{\pi}$&$\wm/2\pi$ (MHz)&$15$  &&\\
&&$\beta$ (ns)& $0.6$ & & \\
\cline{2-6}
&&$\Am$ &$0.45$&& \\
$\tg=24~\ns$ &W-W $\Ra{\hat{n},01}{\pi/2}$&$\wm/2\pi$ (MHz) & $22.5$ &&\\
&&$\beta$ (ns)& $0.7$ & & \\
\cline{2-6}
&DRAG $\Ra{\hat{n},01}{\theta}$&$\beta$ (ns)& $0.6$ & & \\
\cline{2-6}
&DRAG $\Rb{\hat{n},01}{\theta}$&$\beta$ (ns)& $0.7$ & & \\
\hline
\end{tabular}
\label{table:deltas}
\end{table}

\subsection{Randomized benchmarking}
The performance of optimized  pulses was measured using randomized benchmarking (RB). In standard RB, random pairs of $\pi$ and $\pi/2$ pulses are applied, all targeting the same transmon.  The last $\pi/2$ pulse is chosen so that the targeted transmon ideally ends in either $\ket{0}$ or $\ket{1}$. For some sequences, this involves replacing the last $\pi/2$ with identity. The fidelity $\mathcal{F}$ (squared overlap) of the final transmon state to the ideal final state is measured for each RB sequence.  The average $\overline{\mathcal{F}}$ over all RB sequences is then plotted as function of the number of $\pi/2$ pulses, $\Npihalf$.

In alternating RB, two standard RB sequences targeting different transmons are interleaved. Pulses targeting one transmon are applied during the buffer separating pulses targeting  the other.
Calibrated virtual $z$ gates are applied both ways: to $\QA$ following a pulse on $\QB$, and viceversa.

\emph{Pulse randomization.} We briefly describe the pulse randomization procedure used to generate RB sequences. We first create at least $5$ pseudo-random trains of $\pm \pi/2$ pulses around $x$ and $y$ (both signs and axes with equal probability), with enough pulses that the complete RB sequence would span $4~\us$.  Each $\pi/2$ pulse train is  then interleaved with a train of $N_{\pi/2}+1$ Pauli randomization pulses. Each Pauli randomization pulse is taken from the set  $\{I, \Rk{x,01}{\pi}, \Rk{x,01}{-\pi}, \Rk{y,01}{\pi}, \Rk{y,01}{-\pi}\}$ with equal probability $1/5$.  We generate eight randomizations of the Pauli pulse train.  Thus, interleaving the $\pi/2$ and Pauli pulse trains produces at least $5\times8=40$ RB sequences.  After the experiment, we learned that the proper choice of Pauli randomization set would have been
\begin{multline*}
\{I, \Rk{x,01}{\pi}, \Rk{x,01}{-\pi}, \Rk{y,01}{\pi},\\
\Rk{y,01}{-\pi}, \Rk{z,01}{\pi}, \Rk{z,01}{-\pi}\},
\end{multline*}
with probability 1/4 for $I$ and $1/8$ for all others, and $z$-axis $\pi$ rotations replaced with virtual $z$ gates.

\emph{Extraction of average error per computational step}. Following Ref.~\onlinecite{Knill08}, we define a computational step as a pair of $\pi$ and $\pi/2$ pulses including buffers [total step time $\ts=2(\tg+\tb)$].
Our estimate of the average error per computational step, $\EPG$, is obtained from a fit of $\overline{\mathcal{F}}(\Npihalf)$. In the absence of leakage, we expect~\cite{Knill08,Magesan12}
\[
\overline{\mathcal{F}}(\Npihalf)=(1-A)e^{-\alpha\Npihalf}+A,
\]
with $A=1/2$ and $\EPG=\left(1-e^{-\alpha}\right)/2$.  Note that $\overline{\mathcal{F}}(0)=1$ because we correct for readout errors.
This functional form fits very well the  $\overline{\mathcal{F}}(\Npihalf)$ data for standard RB and for alternating RB with
$\QA$ Wah-Wah pulses, with best-fit asymptotic fidelity $A=0.50\pm0.05$ in all cases. $\EPG$ error bars in Figs.~$3$ and \ref{fig:rb20} represent $95\%$ confidence intervals.

In the presence of leakage, we expect~\cite{Epstein13}
\[
\overline{\mathcal{F}}(\Npihalf)=\frac{1}{2}e^{-\alpha\Npihalf}+ \left(\frac{1}{2}-A\right)e^{-\gamma\Npihalf}+A,
\]
with reduced  asymptotic fidelity $1/3\leq A<1/2$.  We find good fits of this form to the $\QB$ $\overline{\mathcal{F}}$ data for alternating RB with $\QA$ DRAG pulses. For $\tg=16~\ns$, $\ts=36~\ns$, the  best-fit $A=0.33\pm 0.02$ confirms strong leakage. Following Ref.~\onlinecite{Epstein13}, we use  $\left(1-e^{-\alpha}\right)/2$ as estimator of $\EPG$ also in this case.

\subsection{Quantum process tomography}
We performed quantum process tomography (QPT) to fully characterize the evolution of $\QB$ under three idling scenarios: no applied pulses on $\QA$, $\QA$ DRAG RB pulsing, and $\QA$ Wah-Wah RB pulsing.

In general, a quantum process is a linear, trace-preserving map of density matrices. The channel can be fully described by a transfer matrix $\PauliTM$ connecting the input and output density matrices, each expanded in a suitable basis. The Pauli basis is a standard choice for processes confined to a qubit subspace~\cite{Chow12}. In our case, the input and output spaces are the qubit and qutrit subspaces of $\QB$. For the input space, we use the basis
\begin{align*}
P^{(2)}_1 & = \begin{pmatrix}1& 0\\0& 0\end{pmatrix},\quad
P^{(2)}_2 = \begin{pmatrix}0& 0\\0& 1\end{pmatrix},\\
P^{(2)}_3 & = \frac{1}{\sqrt{2}} \begin{pmatrix}0& 1\\1& 0\end{pmatrix},\quad
P^{(2)}_4 = \frac{1}{\sqrt{2}} \begin{pmatrix}0& -i\\i& 0\end{pmatrix}.
\end{align*}
For the output space, we use
\begin{align*}
P^{(3)}_1 & = \begin{pmatrix}1& 0 & 0\\0& 0 & 0\\0& 0 & 0\end{pmatrix},\
P^{(3)}_2  = \begin{pmatrix}0& 0 & 0\\0& 1 & 0\\0& 0 & 0\end{pmatrix},\
P^{(3)}_3  = \begin{pmatrix}0& 0 & 0\\0& 0 & 0\\0& 0 & 1\end{pmatrix},\\
P^{(3)}_4 & = \frac{1}{\sqrt{2}}\begin{pmatrix}0& 1 & 0\\1& 0 & 0\\0& 0 & 0\end{pmatrix},\quad
P^{(3)}_5  = \frac{1}{\sqrt{2}}\begin{pmatrix}0& -i & 0\\i& 0 & 0\\0& 0 & 0\end{pmatrix},\\
P^{(3)}_6 & = \frac{1}{\sqrt{2}}\begin{pmatrix}0& 0 & 1\\0& 0 & 0\\1& 0 & 0\end{pmatrix},\quad
P^{(3)}_7  = \frac{1}{\sqrt{2}}\begin{pmatrix}0& 0 & -i\\0& 0 & 0\\i& 0 & 0\end{pmatrix},\\
P^{(3)}_8 & = \frac{1}{\sqrt{2}}\begin{pmatrix}0& 0 & 0\\0& 0 & 1\\0& 1 & 0\end{pmatrix},\quad
P^{(3)}_9  = \frac{1}{\sqrt{2}}\begin{pmatrix}0& 0 & 0\\0& 0 & -i\\0& i & 0\end{pmatrix}.\\
\end{align*}
The quantum process is fully characterized by a $9 \times 4$ real-valued matrix with elements $\PauliTM_{pq}$.

We used a four-step QPT protocol to extract $\PauliTM$ for each idling scenario. The steps are: (i) state preparation, (ii) idling for a time $\ti$, (iii) measurement pre-rotation, and (iv) measurement.

(i) For state preparation, a calibrated DRAG pulse $U_{n}$ was applied to $\ketB{0}$, taken from the set
\begin{multline*}
\{I, \Rb{x,01}{\pi}, \Rb{x,01}{\pi/2}, \Rb{x,01}{-\pi/2},\\
\Rb{y,01}{\pi/2}, \Rb{y,01}{-\pi/2}\}.
\end{multline*}

(ii) The $\QA$ rotations during $\ti$ were chosen according to the RB protocol described above. The QPT protocol was repeated for $64$ distinct RB sequences (8 seeds, 8 Pauli randomizations per seed). The transfer matrix $\PauliTM$ was computed as an average over these randomizations.

(iii) The measurement pre-rotations $V_m$ on $\QB$ were chosen from the set
\begin{multline*}
\{I,\Rb{x,01}{\pi/2},\Rb{x,01}{-\pi/2},\Rb{y,01}{\pi/2},\Rb{y,01}{-\pi/2},\\
\Rb{x,01}{\pi},\Rb{x,12}{\pi/2},\Rb{x,12}{-\pi/2},\\
\Rb{y,12}{\pi/2},\Rb{y,12}{-\pi/2},\Rb{x,01}{\pi}\Rb{x,12}{\pi/2},\\
\Rb{x,01}{\pi}\Rb{x,12}{-\pi/2}, \Rb{x,01}{\pi}\Rb{y,12}{\pi/2},\\
\Rb{x,01}{\pi}\Rb{y,12}{-\pi/2},\Rb{x,01}{\pi}\Rb{x,12}{\pi}\}.
\end{multline*}
Optimized DRAG pulses were used to implement all rotations.  This set of pre-rotations augments that of Ref.~\cite{Bianchetti10} with redundant rotations in order to increase the stability of the inversion.

(iv) Using the dispersive readout described above, we obtain averaged integrated voltages $\langle V_I \rangle$ and $\langle V_Q \rangle$ for each $(V_m, U_n)$ pair.

The averaged measurement $\mVi_{kl}$ for each $(V_m,U_n)$ pair ($6\times15=90$ pairs total) is related to $\PauliTM$ by
\begin{equation}
\mVi_{mn} = \sum_{pq} \PauliTM_{pq} \tr (V_m^\dag M_i V_m P^{(3)}_p) \bra{0} U_n^\dag P^{(2)}_q U_n \ket{0}.\label{Eq:mikl}
\end{equation}
Combining all measurements, we arrive at a set of 180 linear equations for the 36 unknown $\PauliTM_{pq}$. We solve this over-determined set of linear equations by unweighted least-squares inversion.

As a measure of idling performance, we extract the average gate fidelity, $\Fgate$, of $\QB$ to identity~\cite{Gilchrist05,Motzoi09}. We find
\begin{align*}
\Fgate & = \sum_{j = \pm x, \pm y, \pm z} \tr\left[\rho_j \mathcal{E} (\rho_j) \right]\\
 & = \frac{\PauliTM_{11} + \PauliTM_{22} + \PauliTM_{43} + \PauliTM_{54} - \PauliTM_{31} - \PauliTM_{32}}{6} + \frac{1}{3}.
\end{align*}
To gain further insight into the sources of infidelity, we decompose $\Fgate$ as
\begin{equation}
\Fgate = \frac{F_1 + F_2 + 1}{3},
\label{eq:fgate}
\end{equation}
where $F_1 = \frac{1}{2}\left(\PauliTM_{11} + \PauliTM_{22} - \PauliTM_{31} -\PauliTM_{32}\right)$ and $F_2 = \frac{1}{2}\left(\PauliTM_{43} + \PauliTM_{54}\right)$. $F_1$ is sensitive to errors in population transfer. $F_2$ is sensitive to population transfer and also to pure dephasing within the qubit subspace.
For true idling, we model
\begin{equation}
F_1(\ti)=\frac{1+e^{-\ti/T_{1(1\rightarrow0)}}}{2},
\label{eq:F1}
\end{equation}
and
\begin{equation}
F_2(\ti)=e^{-\ti/(2T_{1(1\rightarrow0)})}e^{-\ti^2/T_\varphi^2}.
\label{eq:F2}
\end{equation}
The model $F_2$ reflects dominant pure dephasing by $1/f$ flux noise, as suggested by a non-exponential Ramsey fringe decay observed for $\QB$.

\section{Simulation}
We perform a numerical simulation of the driven two-qubit system in order to: (a) validate the Wah-Wah modulation parameters  manually optimized at bias point 1 and (b)  speed-up the optimization of these parameters
at other bias points. Following Ref.~\onlinecite{Schutjens13}, we model the system Hamiltonian in a frame rotating with a resonant drive at $\wA$, truncate at three lowest-energy levels per transmon, and make the rotating
wave approximation:
\begin{eqnarray}
\label{eq:h}
\hat{H}/\hbar &&= \DA \hat{\Pi}_2^a + (\delta-\DB)\hat{\Pi}_1^b+\delta \hat{\Pi}_2^b \nonumber\\
&&+\frac{\Omega_I(t)}{2}\left[ \lambda_1^a \hat{\sigma}_{x,1}^{a} + \lambda_1^b \hat{\sigma}_{x,1}^{b}+ \lambda_2^a \hat{\sigma}_{x,2}^{a} + \lambda_2^b \hat{\sigma}_{x,2}^{b}\right]\nonumber\\
&&+\frac{\Omega_Q(t)}{2}\left[ \lambda_1^a \hat{\sigma}_{y,1}^{a} + \lambda_1^b \hat{\sigma}_{y,1}^{b}+ \lambda_2^a \hat{\sigma}_{y,2}^{a} + \lambda_2^b \hat{\sigma}_{y,2}^{b}\right].\nonumber\\
\end{eqnarray}
Here, $\sigma_{x,l}^{k}=\ket{l_k}\bra{l-1_k}+\ket{l-1_k}\bra{l_k}$, and
$\sigma_{y,l}^{k}=i\ket{l_k}\bra{l-1_k}-i\ket{l-1_k}\bra{l_k}$. As defined in the main text, $\Delta_k=\omega_{12}^{k}-\omega_{01}^{k}$, $\delta=\omega_{12}^{b}-\omega_{01}^{a}$, and $\Omega_I(t)$ and $\Omega_Q(t)$ are the in- and out-of-phase pulse envelopes. Finally, $\lambda_l^{k}$ is the coupling strength of the drive to the $\ket{l-1_k}\leftrightarrow\ket{l_k}$ transition. All model parameters are obtained from calibration measurements.

For each choice of gate time $\tg$, bias point, and $\QA$ pulse rotation angle $\theta \in \{ \pi, \pi/2 \}$, we identify a manifold of $(\Atheta,\beta,\Am,\wm)$ values performing a high-quality pulse on $\QA$. For each $(\Am,\wm)$ pair in the range $\Am\in[-1,1]$ and $\wm/2 \pi \in [0,100~\MHz]$, we find the $\Atheta$ and $\beta$ achieving the desired rotation angle $\theta$ and minimizing internal leakage in $\QA$. We then calculate the leakage $\Ptwob$ induced by four back-to-back $\QA$ pulses, starting from $\left(\ketB{0}+i\ketB{1}\right)/\sqrt{2}$. Similarly to Fig.~2, the phase of these $\QA$ pulses is increased in progression ($\phi$, $2 \phi$, $3 \phi$, and $4 \phi$).  We repeat for 200 values of $\phi$ between 0 and $2\pi$. The simulation output consists of an image plot of $\max_{\phi} \Ptwob$ as a function of $\Am$ and $\wm$.

\section{Extended Results}
\begin{figure}
\includegraphics{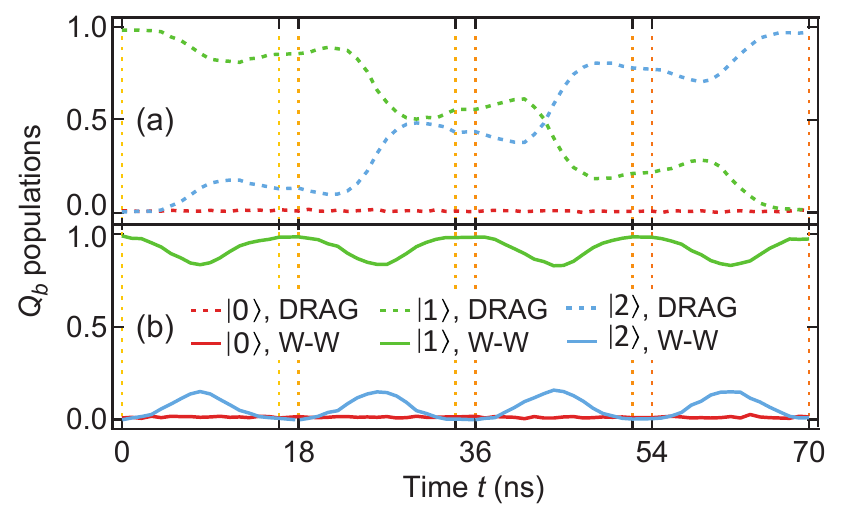}
\caption{Measured evolution of $\QB$ level populations during four back-to-back $\pi/2$ pulses on $\QA$, with optimized DRAG and Wah-Wah envelopes ($\tg=16~\ns$). Panels (a) and (b) are dual to Figs.~2(c) and 2(d), respectively, but with $\pi/2$ instead of $\pi$ pulses on $\QA$. The phase of $\pi/2$ pulses is $\phi$, $2\phi$, $3\phi$, and $4\phi$, in progression, with $\phi=319^\circ$.
}
\label{fig:strobopi2}
\end{figure}

\begin{figure*}
\includegraphics{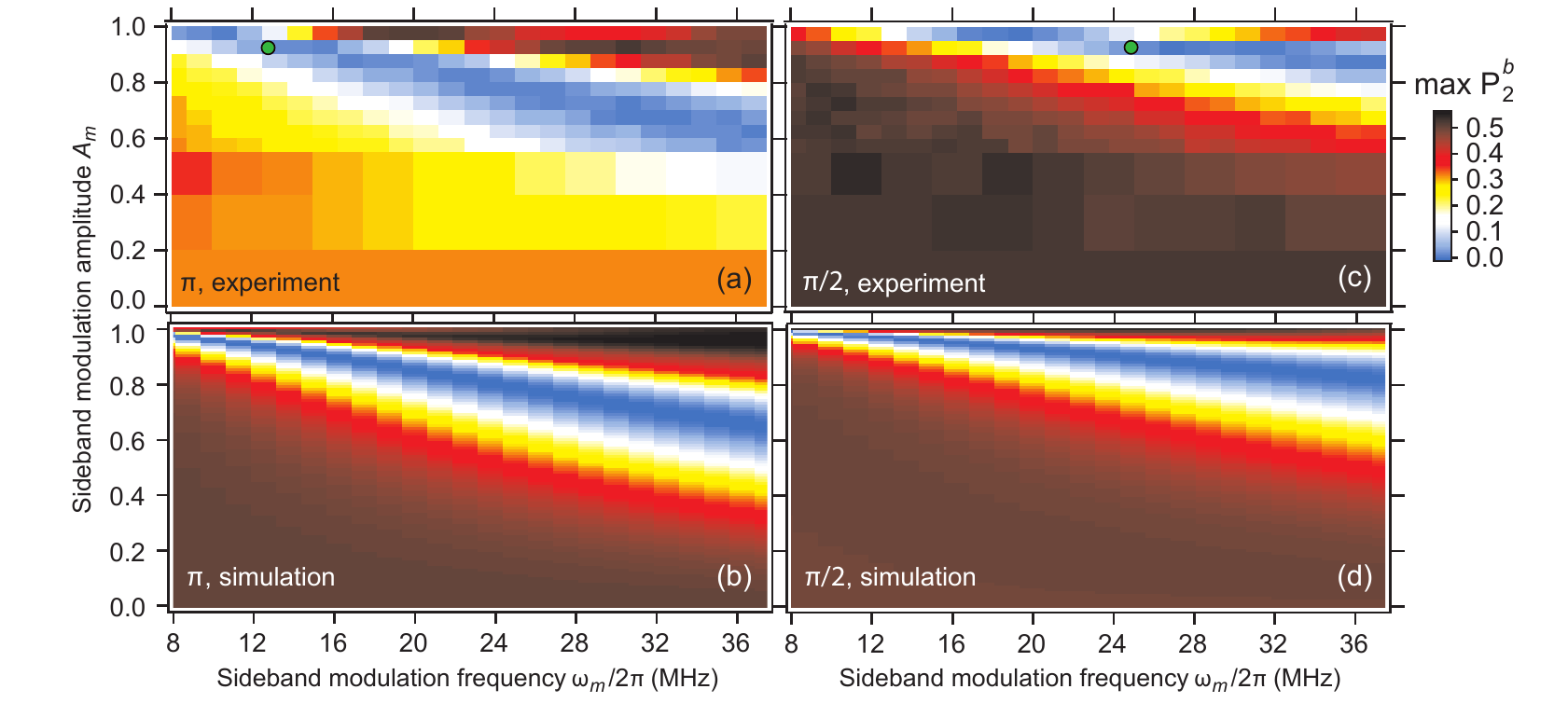}
\caption{Image plots of measured (a,c) and simulated (b,d) final population of $\ketB{2}$, $\Ptwob$, as a function of the  modulation parameters $\Am$ and $\wm$ in four consecutive $\pi$ (a,b) or $\pi/2$ (c,d) $\QA$ Wah-Wah pulses ($\tg=16~\ns$). Transmons are initially in $\ketA{0}$ and $\left(\ketB{0}+i\ketB{1}\right)/\sqrt{2}$. Similarly to Figs.~2 and \ref{fig:strobopi2}, the phases of the pulses are $\phi$, $2 \phi$, $3 \phi$, and $4 \phi$, in progression.  For each $(\Am,\wm)$ pair, we plot the maximal $\Ptwob$ measured over $80$ values of  $\phi$ between 0 and $2\pi$. In experiment (a,c), we optimized the $\beta$ coefficient at the left and right boundaries, and used linear interpolation with respect to $\wm$ (at fixed $\Am$) to set $\beta$ inside.  Markers indicate the manually found $(\Am,\wm)$ pairs minimizing $\Ptwob$ in experiment.
}
\label{fig:simuboth}
\end{figure*}

\begin{figure*}
\includegraphics{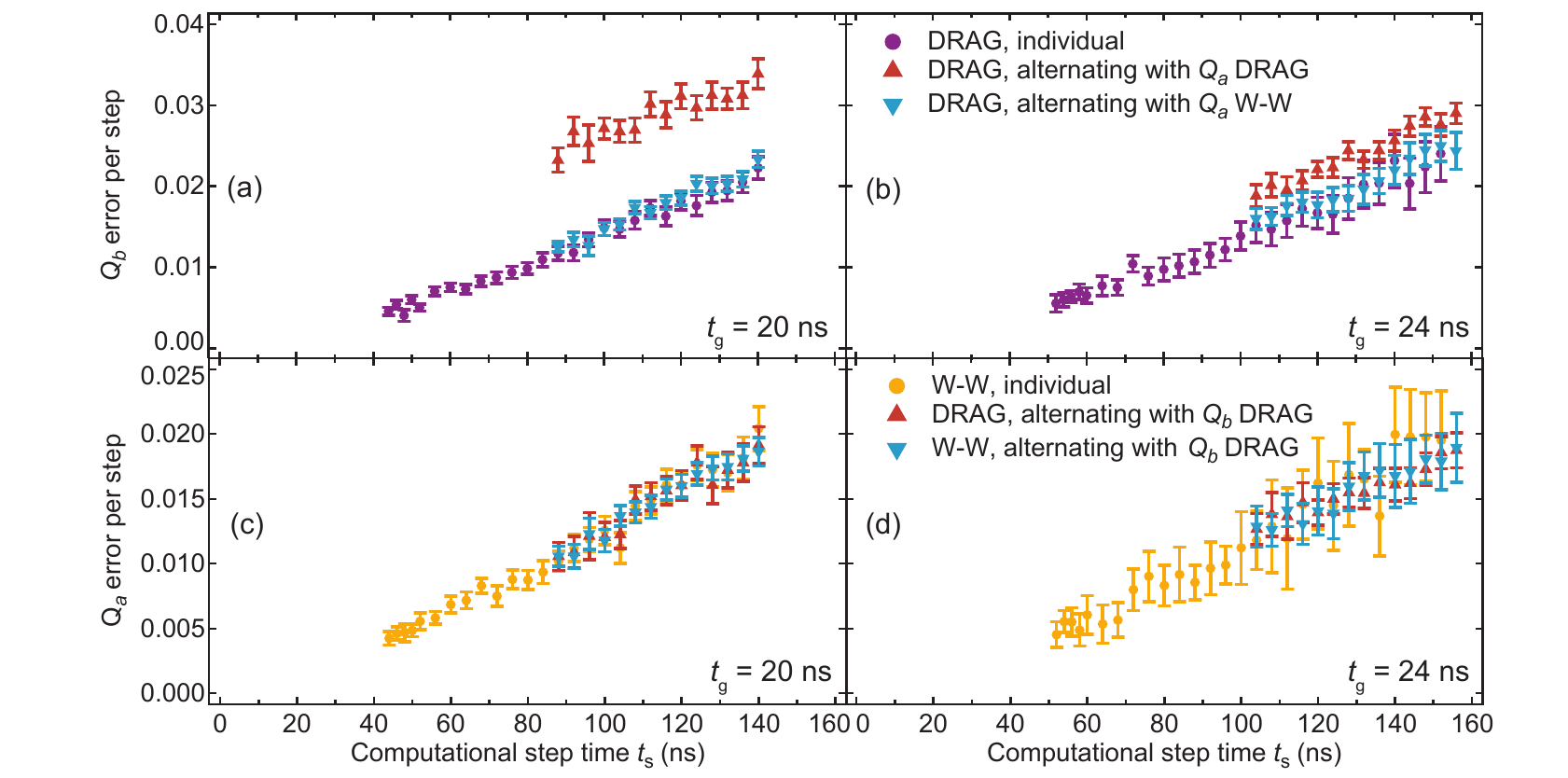}
\caption{Error per computational step, $\EPG$, as a function of the computational step time $\ts=2(\tg +\tb)$, obtained in standard and alternating randomized benchmarking. This figure is dual to Fig.~3, but with gate times $\tg=20~\ns$ (a,c) and $\tg=24~\ns$ (b,d), corresponding to $\tg |\delta|/2\pi \approx 1.1$ and $1.4$, respectively. As expected, the $\QB$ leakage induced by $\QA$ DRAG pulses is less severe the longer $\tg$. Optimized $\QA$ Wah-Wah pulsing allows achieving decoherence-limited $\EPG$ on $\QB$.}
\label{fig:rb20}
\end{figure*}

\begin{figure}
\includegraphics{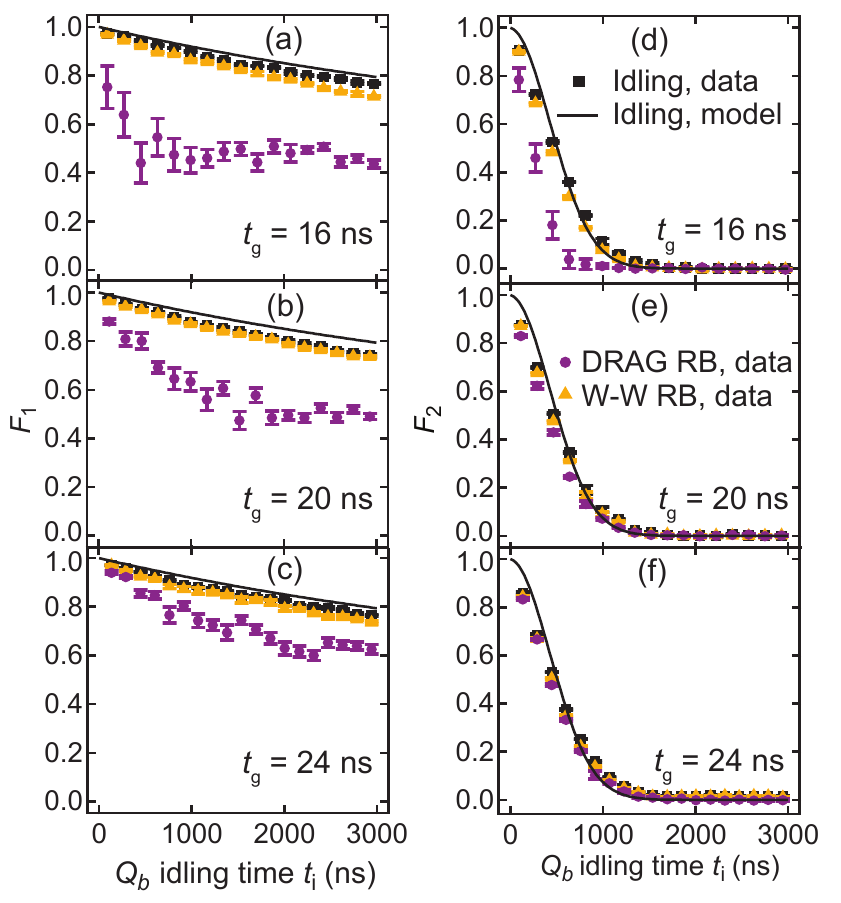}
\caption{
$\QB$ average gate fidelity to the identity process as a function of idling time $\ti$, decomposed into components with different sensitivity to population transfer and to pure dephasing. (a-c) Fidelity $F_1$ is sensitive to relaxation within the qubit subspace and leakage out of the qubit subspace.  (d-f) Fidelity $F_2$ is also sensitive to pure dephasing within the qubit subspace. Same raw data as in Fig.~4. Squares, circles, and triangles correspond to true idling (i.e., no pulses on $\QA$), $\QA$ DRAG RB, and $\QA$ Wah-Wah RB, respectively. The combination of these plots allows two conclusions: $\QB$ idling for $\QA$ Wah-Wah RB is decoherence limited; $\QB$ idling for $\QA$ DRAG RB is dominated by leakage (rather than imperfect $z$-gate compensation). The large error bars for $\QA$ DRAG RB reflect the sensitivity of $\QB$ leakage to the randomization of $\QA$ DRAG pulses. See text for the details of the model for true idling (solid curves).
}
\label{fig:fid12}
\end{figure}

This section presents four figures lending further support to the main text claims. Complementing Fig.~2, Fig.~\ref{fig:strobopi2} shows the evolution of level populations in $\QA$ and $\QB$ during repeated $\pi/2$ pulses on $\QA$. Figure~\ref{fig:simuboth} compares measurements and simulation of $\ketB{2}$ leakage induced by four back-to-back $\pi$ or $\pi/2$ Wah-Wah pulses on $\QA$. Image plots of leakage as a function of modulation parameters show good correspondence between the optimal Wah-Wah parameters found by simulation and manually in experiment. Complementing Fig.~3, Fig.~\ref{fig:rb20} shows standard and alternating randomized benchmarking results for longer gate times $\tg=20~\ns$ and $24~\ns$. Finally, complementing Fig.~4, Fig.~\ref{fig:fid12} helps identify dominant limitations to $\QB$ idling from decoherence, leakage error, and phase-compensation error.

\bibliographystyle{prl}
\bibliography{../../../TeX/References/References_cQED}